\documentclass[10pt,letterpaper]{article}
\usepackage[top=0.85in,left=2.75in,footskip=0.75in,marginparwidth=2in]{geometry}

\usepackage[utf8]{inputenc}

\usepackage{cite}

\usepackage{nameref,hyperref}
\usepackage{booktabs}
\usepackage[right]{lineno}
\usepackage{amsmath}
\usepackage{microtype}
\DisableLigatures[f]{encoding = *, family = * }

\raggedright
\usepackage{changepage}

\usepackage[aboveskip=1pt,labelfont=bf,labelsep=period,singlelinecheck=off]{caption}

\makeatletter
\renewcommand{\@biblabel}[1]{\quad#1.}
\makeatother

\usepackage{lastpage,fancyhdr,graphicx}
\usepackage{epstopdf}
\fancyheadoffset[L]{2.25in}
\fancyfootoffset[L]{2.25in}

\usepackage{color}

\definecolor{Gray}{gray}{.25}

\usepackage{graphicx}

\usepackage{sidecap}

\usepackage{wrapfig}
\usepackage[pscoord]{eso-pic}
\usepackage[fulladjust]{marginnote}
\reversemarginpar

\begin{document}
\vspace*{0.35in}

% title goes here:
\begin{flushleft}
{\Large
\textbf\newline{Parallel tempering as a mechanism for facilitating inference in hierarchical hidden Markov models}
}
\newline
% authors go here:
\\
Giada Sacchi\textsuperscript{1,2},
Ben Swallow\textsuperscript{1,*}
\\
\bigskip
\bf{1} School of Mathematics and Statistics, University of Glasgow, UK
\\
\bf{2} School of Mathematics and Statistics, University of Edinburgh, UK
\\
\bigskip
* ben.swallow@glasgow.ac.uk

\end{flushleft}

\section*{Abstract}
The study of animal behavioural states inferred through hidden Markov models and similar state switching models has seen a significant increase in popularity in recent years. The ability to account for varying levels of behavioural scale has become possible through hierarchical hidden Markov models, but additional levels lead to higher complexity and increased correlation between model components. Maximum likelihood approaches to inference using the EM algorithm and direct optimisation of likelihoods are more frequently used, with Bayesian approaches being less favoured due to computational demands. Given these demands, it is vital that efficient estimation algorithms are developed when Bayesian methods are preferred. We study the use of various approaches to improve convergence times and mixing in Markov chain Monte Carlo methods applied to hierarchical hidden Markov models, including parallel tempering as an inference facilitation mechanism. The method shows promise for analysing complex stochastic models with high levels of correlation between components, but our results show that it requires careful tuning in order to maximise that potential.

% the * after section prevents numbering

\section{Introduction}

In a crucial moment for the environment, while climate is rapidly changing and putting flora and fauna under great threat, it is important to investigate the evolution of ecological populations, in order to estimate demographic parameters and forecast their future developments.

The study of individuals of the same species and their behaviours, how they constitute the populations in which they exist and how such populations evolve is called \textit{population ecology} \cite{king}.
Behavioural models are a mathematical tool used to describe the behaviour of an entity or a phenomenon. 
In particular, they can be employed in population ecology to reflect the patterns in animal behaviours and movements, enabling the analysis and identification of different modes over time.

Hidden Markov Models (HMMs) and associated state-switching models are becoming increasingly common time series models in ecology, since they can be used to model animal movement data and infer various aspects of animal behaviour \cite{Leos1}. 
Indeed, they are able to model the propensity to persist in such behaviours over time and to explain the serial dependence typically found, by enabling the connection of observed data points to different underlying ecological processes and behavioural modes. 

Several publications in the field of statistical ecology have benefited from the versatility of HMMs for the analysis of animal behaviours \cite{diecks}, focusing especially on dissecting movement patterns of single individuals into different behavioural states \cite{langrock}.
In fact, HMMs have a great interpretive potential that allows to deal with unmeasured state processes and identify transitions in ``hidden'' states, even if such transitions are not evident from the observations \cite{tucker}.
By formally extricating state and observation processes based on manageable yet powerful mathematical properties, HMMs can be used to interpret many ecological phenomena, facilitating inferences about complex system state dynamics that would otherwise be intractable \cite{langrock20}.
Thus, the success of applying HMMs in ecological systems lies in the combination of biological expertise with the use of sophisticated movement models as generating mechanisms for the observed data \cite{Leos2}.

Nevertheless, many extensions of basic HMMs have been investigated only recently in statistical literature, implying that HMMs for modelling animal movement data have not been recognised yet to their full extent in ecological applications \cite{langrock}. For example, the so-called \textit{hierarchical} HMMs used in this project have been already employed in order to distinguish between different handwritten letters, but also to recognise a word, defined as a sequence of letters \cite{fine01}. The framework proposed in \cite{Leos1} paves the way to the application of such an extension of Hidden Markov Models for the simultaneous modelling of animal behaviour at distinct temporal scales, in the light of the idea that adding hierarchical structures to the HMM opens new possibilities in the sphere of animal movement and behaviour inference.

Considering a sequence of latent production states -and associated observations- as the manifestation of some behavioural processes at a cruder temporal scale, the concept of HMM can be extended to build a hierarchical process with multiple time scales (henceforth referred to as \textit{multi-scale behaviours}). By jointly modelling multiple data streams at different temporal resolutions, corresponding models may help to draw a much more comprehensive picture of an animal's movement patterns, for example with regard to long-term versus short-term movement strategies \cite{langrock}.

The modelling framework used in this project was proposed and applied to the same collection of data by \cite{Leos1}. Data were assumed to stem from two behavioural processes, operating on distinct temporal scales: a crude-scale process that identifies the general behavioural mode (e.g. migration), and a fine-scale process that captures the behavioural mode nested within the large-scale mode (e.g. resting, foraging, traveling). 
Intuitively, the former may persist for numerous consecutive dives, whereas the latter agrees to the more nuanced state transitions at the dive-by-dive level, given the general behavioural mode \cite{Leos1}. 
Hence, a behaviour occurring at the crude time scale can be connected to one of the finite internal states, such that each internal state generates a distinct HMM, the internal states of which in turn are linked to the actual observation at a specific point in time. 

This paper aims to devise a successful strategy for the analysis of movement data related to a harbour porpoise through a Bayesian approach, dealing with the correlations frequently inherent between the varying processes \cite{touloupou} through the use of parallel inference schemes that can avoid associated computational issues.

Although it follows the modelling structure of \cite{Leos1}, the two studies differ in the type of inference method adopted. In fact, the main purpose of this research is constructing a Bayesian framework for the stochastic problem addressed in \cite{Leos1}, as opposed to the frequentist investigation therein.
Various MCMC algorithms will be implemented on the basis of requirements posed by either the data or the design of the study, with the conclusions reached through the Bayesian approach compared to the previous results.

\section{Materials and Methods}

\subsection{Basic Hidden Markov Models}
\label{sec:hmm1}

A basic Hidden Markov Model is a doubly stochastic process with observable state-dependent processes $\{Y_t\}^T_{t=1}$ controlled by underlying state processes $\{S_t\}^T_{t=1}$, through the so-called \textit{state-dependent distributions} $\{f_n\}^N_{n=1}$ \cite{Leos2}.

Under the assumption that the latent \textit{production state} at time $t$, $S_t$, can take on a finite number $N \geq 1$ of states as generated by the corresponding $f_n$, the evolution of states over time is governed by a Markov chain with t.p.m. $\Gamma^{(t)}=\gamma^{(t)}_{i,j}$, where $\gamma^{(t)}_{i,j}=Pr(S_t=j|S_{t-1}=i)$ for $i,j=1,...,N$.
Hence, the distribution of the production state $S_t$ is fully determined by the previous state $S_{t-1}$.

It should be further assumed that any observation $Y_t$ is conditionally independent of past and future observations and production states, given the current production state $S_t$. In this way, the production states effectively select which of the finitely many possible distributions each observation belongs. 

The state-dependent distributions for $Y_t$ can be represented in terms of conditional probability density (or mass) functions $f(y_t|S_t = i) = f_i(y_t), \text{ for } i = 1,...,N$, often also called \textit{emission probabilities} \cite{yoon}. For multivariate observations, in which case $Y_t = (Y_{1t},...,Y_{Rt})$, the options are to either formulate a joint distribution $f_i(y_t)$ or assume contemporaneous conditional independence by allowing the joint distribution to be represented as a product of marginal densities, $f_i(y_y)=f_i^1(y_{1t}) f_i^2(y_{2t}) ... f_i^R(y_{Rt})$ \cite{Leos1}. 

Assuming a time-homogeneous process, then $\Gamma^{(t)}=\Gamma$.  
Furthermore, defining the initial state distribution $\pi^{(1)}$ as the vector with entries $\pi_n^{(1)}=Pr(S_1=n)$, for $n=1,...,N$, $\pi^{(1)}$ is usually taken to be the stationary distribution, solving $\Gamma\pi=\Gamma$ \cite{Leos2}. 

The likelihood of the observations can be obtained by marginalizing their joint distributions. In a basic HMM, this requires the summation over all possible production state sequences $s_t$,
\begin{equation} \label{eq:lk0}
L_p=\sum_{s_t=1}^N\sum_{s_t=1}^N\pi_{s_1}\prod_{t=2}^T\gamma{s_{t-1},s_t}\prod_{t=1}^T f_{s_t}(y_t).
\end{equation}
According to \cite{zucchini}, Eq. \ref{eq:lk0} can be also written explicitly as a matrix product,
\begin{equation} \label{eq:lk1}
L_p=\pi^T Pr(y_1)\Gamma Pr(y_2)...\Gamma Pr(y_T)\textbf{1}^T
=\pi^T Pr(y_1)\prod_{t=2}^T\Gamma Pr(y_t)\textbf{1}^T,
\end{equation}
for an initial distribution $\pi$, a $N \times N$ matrix $Pr(y_t) = diag (f_1(y_t), ... , f_N (y_t))$, and a column vector of 1's $\textbf{1}^T$. \cite{langrock}.

\subsection{Multiple Hidden Markov Models}
\label{sec:hmm2}

For a hierarchical Hidden Markov Model, two types of latent states (production states and internal states) are assumed to occur at different temporal scales. The production state $S_t$ is taken to be generated depending on which of $K$ possible  internal states is active during the current time frame.
Therefore, a fine-scale sequence of observations, $y_m = (y_{1,m}, ... , y_{T,m})$ -with one such sequence for each $m = 1, ... , M$- can be thought as produced by a sequence of production states, $S_{1,m}, ... , S_{T,m}$ during a given time frame (namely the $m$-th of $M$ time frames) \cite{Leos1}. 

The length of the sequence of the production states produced by the $k$-th internal state can be chosen according to the data collection process or dictated by the analysis. The corresponding $K$-state internal state process, $\{H_m\}^M_{m=1}$, is such that $H_m$ serves as a proxy for a behaviour occurring at a cruder time scale, namely throughout the $m$-th time frame \cite{Leos2}. 

Assuming a Markov chain at the time frame level, then the $m$-th internal state is conditionally independent of all the others given the internal state at the $(m-1)$-th time point, i.e. $Pr(H_m|H_{m-1}, ..., H_1) = Pr(H_m|H_{m-1})$. Thus, the $K \times K$ t.p.m. for the internal states $\{H_m\}^M_{m=1}$ examines both the persistence in the internal states and the transitioning between them \cite{Leos1}. 

Defining the likelihood for the production states as in Eq. \ref{eq:lk1}, the conditional likelihood for a production state given the $k$-th internal state can be written as $L_p(y_m|H_m=k)$, for $k=1,...,K$. Denote by $\pi^{(I)}$ a $K$-vector of initial probabilities for the internal states and by $\Gamma^{(I)}$ the $K \times K$ t.p.m. for the internal state process. 
Then, the marginal likelihood for a hierarchically structured HMM can be expressed as
\begin{equation}
    L_{p,m}=\pi^{(I)}Pr^{(I)}(y_1)\prod_{m=2}^M \Gamma^{(I)}Pr^{(I)}(y_m)\textbf{1}^T,
\end{equation}
where $Pr^{(I)}(y_m) = diag (L_p(y_m|H_m=1), ... , L_p(y_m|H_m=K))$ is a $K \times K$ matrix \cite{Leos1}.

\subsection{Bayesian Inference}
\label{sec:bayesinf}

Markov chain Monte Carlo (MCMC) methods are simulation techniques that perform Monte Carlo integration using a Markov chain to generate observations from a target distribution. 
The Markov chain is constructed in such a way that, as the parameters are updated -and the number of iterations $t$ increases-, the distribution associated with the $t$-th observation gets closer to the target distribution. 
Once the chain has converged to the stationary distribution $\pi$, the sequence of values taken by the chain can be used to obtain empirical (Monte Carlo) estimates of any posterior summary, including posterior distributions for parameters and latent variables.
Therefore, MCMC algorithms are designed so that the Markov chain converges to the joint posterior distribution of the parameters given the model, the data, and prior distributions. Many different samplers have been created to construct Markov chains with the desired convergence properties and the efficiency of the resulting sample generation can vary drastically depending on the choice of algorithm and the tuning of parameters there-in.

The Gibbs sampler is an MCMC algorithm for generating random variables from a (marginal) distribution directly, without having to calculate the density. It is useful when the joint distribution of the parameters is not known explicitly or is difficult to sample from directly, but it is feasible to sample from the conditional distribution of each parameter. Given the state of the Markov chain at the $t$-th iteration, the Gibbs sampler successively draws randomly from the full posterior conditional distributions of $\theta_p$.

The Metropolis-Hastings algorithm is a MCMC method for generating a sequence of states from a probability distribution from which it would be difficult to sample directly. 
The principle of the algorithm is to sample from a \textit{proposal \emph{(or} candidate) distribution} $q$, which is a crude approximation of the (posterior) target distribution $h$ \cite{handbook}. Observations are subsequently drawn from the proposal distribution, conditional only upon the last observation. This induces a Markov process that asymptotically reaches the unique stationary distribution $\pi(\theta)$, such that $\pi(\theta)=h(\theta|x)$. The Metropolis–Hastings algorithm can be illustrated as follows.
\begin{enumerate}
    \item[I.] Initialise
    \begin{enumerate}
        \item[1.] Set an initial state $\theta^1$
    \end{enumerate}
    \item[II.]  Iterate (for $t=1,2,...$)
    \begin{enumerate}
        \item[1.] Generate a random candidate state $\theta^*$ according to the proposal distribution $q(\theta^*|\theta^t)$
        \item[2.] Calculate the acceptance probability $\alpha(\theta^t, \theta^*)$
        \item[3.] Generate a uniform random number $u\in[0,1]$
            \begin{itemize}
            \item if $u\leq \alpha(\theta^*,\theta^t)$, then accept the proposed move and set $\theta^{t+1}= \theta^*$
            \item if $u > \alpha(\theta^*,\theta^t)$, then reject the proposed move and set $\theta^{t+1}= \theta^t$
            \end{itemize}
        \end{enumerate}
\end{enumerate}

\subsection{Parallel Tempering}
\label{sec:pt}

MH algorithms might get stuck in local maxima, especially in high-dimensional problems or multimodal densities.
A feasible solution is trying to run a population of Markov chains in parallel, each with possibly different, but related stationary distributions. Information exchange between distinct chains enables the target chains to learn from past samples, improving the convergence to the target chain \cite{gupta}.

\textit{Parallel tempering} (PT) is a method that attempts periodic swaps between several Markov chains running in parallel at different temperatures, in order to accelerate sampling. Each chain is equipped with an invariant distribution connected to an auxiliary variable, the \textit{temperature} $\beta$, which scales the ``shallowness'' of the energy landscape, and hence defines the probability of accepting an unsuitable move \cite{gupta,hansmann}. Following \cite{metropolis53}, the energy of a parameter $\theta$ is defined as $E[\theta]=-logL[\theta]-logp[\theta]$, where $L$ and $p$ are the likelihood and the prior probability, respectively.

The algorithm is derived from the idea that an increase in the temperature smooths the energy landscape of the distribution, easing the MH traversal of the sample space. In fact, high temperature chains produce high density samples that accept unfavourable moves with a higher probability, thus exploring the sample space more broadly. As a result, switching to higher temperature chains allows the current sampling chain to circumvent local minima and to improve both convergence and sampling efficiency \cite{chib}.

The Parallel Tempering algorithm for a parameter vector $\theta$ can be illustrated as follows.
\begin{itemize}
    \item For $s=1,...,S$ swap attempts
    \begin{enumerate}
        \item[1.] For $j=1,...,J$ chains
        \begin{enumerate}
            \item[I.] For $t=1,...,T_{MCMC}$ iterations
            \begin{enumerate}
                \item[i] Propose a new parameter vector $\theta^*$
                \item[ii] Calculate the energy $E[\theta^*]$
                \item[iii] Set $\theta_{t+1}=\theta^*$ with probability $min(1, e^{-\beta_j \Delta E_j})$, where $\Delta E_j=E_j[\theta^*]- E_j[\theta_{t-1}].$ Otherwise, set $\theta_{t+1}=\theta_t$.
            \end{enumerate}
            \item[II.] Record the value of the parameters and the energy on the final iteration
        \end{enumerate}
        \item[2.] For each consecutive pair of chains (in decreasing order of temperature)
        \begin{itemize}
            \item[I.] Accept swaps with probability $min(1,e^{-\Delta \beta \Delta E})$, where $\Delta E=E_j-E_{j-1}$ and $\Delta \beta= \beta_j-\beta_{j-1}.$
        \end{itemize}
    \end{enumerate}
\end{itemize}

\subsection{Data}

The data relate to the harbour porpoise (\textit{Phocoena phocoena}) dives presented in \cite{Leos1}.
So as to build the multi-scale hierarchical structure for the Hidden Markov Model, dive patterns were inferred at a crude scale $K$-state Markov process that identifies the general behavioural mode, and at a fine scale for state transitions at a dive-by-dive resolution, given the general behaviour.
Hence, a behaviour occurring at a crude time scale can be connected to one of the $K$ internal states, such that each internal state generates a distinct HMM, with the corresponding $N$ production states generating the actual observation at a specific point in time \cite{Leos1}.
The production states are thus used to identify and categorize behavioural states, whilst internal states analyse patterns of changes in such behavioural states.

The crude time scale, hereafter also referred to as \textit{lower level}, was built based on the notion that a dive pattern is typically adopted for several hours before switching to another one. 
For this purpose, observations recorded per second were grouped into hourly intervals, allowing each segment to be connected to one of the $K=2$ HMMs with $N=3$ (dive-by-dive \textit{upper level}) states each. 

Across the two dive-level HMMs, the same state-dependent distributions but different t.p.m.s were employed. 
This implied that any of the three types of dive (originating each from one of the three different production states) could occur in both lower level behavioural modes, but should not manifest equally often, on average, due to the different Markov chains active at the upper level. 
Moreover, supposing $M$ time frames per individual, differences observed across $y_m$, for $m = 1, ... , M$, were explained by considering the way in which a porpoise switches among the $K$ internal states, while still modelling the transitions among production states at the time scale at which the data were collected \cite{Leos1}.

The initial state distributions, for both the internal and the production state processes, were assumed to be the stationary distributions of the respective Markov chains. 

The raw data had been processed using the R package \textsc{diveMove} and transformed into measures of \textit{dive duration, maximum depth \emph{and} dive wiggliness} to characterize these Cetaceans' vertical movements at a dive-by-dive resolution, where \textit{dive wiggliness} refers to the absolute vertical distance covered at the bottom of each dive \cite{Leos1}.
Such measures were considered to be the covariates characterizing the model at the upper level, consisting of the observable fine-scale sequences $y_m = (y_{1,m}, y_{2,m}, ... , y_{T,m})$ for $m = 1, ... , M$.

Furthermore, in order to construct a relatively simple yet biologically informative model, gamma distributions and contemporaneous conditional independence were assumed for the three covariates. Hence, for any given dive, the observed variables were considered conditionally independent, given the production state active at the time of the dive \cite{Leos1}.

\subsection{Parameters}
\label{sec:par}
The parameters to be estimated through Bayesian inference are the ones defining the gamma distributions of the covariates for all production states, with an additional point mass on zero for the \textit{dive wiggliness} to account for null observations.

In principle, covariates could be included both in the state-dependent processes, where they determine the parameters of the state-dependent distributions, and in the state processes, where they affect the transition probabilities \cite{langrock}. 
However, in ecology the focus is usually on the latter, as the interest lies in modelling the effect of covariates on state occupancy.
Therefore, it is of interest to estimate also the transition probabilities matrices $\Gamma$ (see {Sect.} \ref{sec:hmm1}), and the corresponding stationary distributions, for all the production states. 
The probabilities $\gamma_{ij}$ at time $t$ can be expressed as a function of some predictor $\eta_{ij}(x_t)$ that in turn depends on a $Q$-dimensional covariate vector $x_t=(x_{1,t}, ..., x_{Q,t})$ \cite{langrock}, where $Q=3$.

In order to ensure identifiability when estimating the entries of $\Gamma$, the ``natural'' parameters $\gamma_{ij}$  were transformed into ``working'' ones, $\eta_{ij}$ \cite{Leos1}.
The $\gamma_{ij}$ values were mapped by row onto the real line with the use of the multinomial logit link, to guarantee that $0<\gamma_{ij}(x_t)<1$ and $\sum_{j=1}^N\gamma_{ij}(x_t)$ for $\forall i$, and the diagonal entries of the matrix were taken as the reference categories:
\[
\gamma_{ij}=\frac{\exp(\eta_{ij})}{\sum_{k=1}^{N}\exp(\eta_{ij})},  \textrm{ where } \eta_{ij}=\begin{cases}
\beta_{ij} & \textrm{if } i \neq j;\\
0 & \textrm{otherwise}.
\end{cases}
\]
For the sake of interpretability, the working parameters were eventually translated to their 

\noindent corresponding natural values.

However, since HMMs lie within the class of mixture models, the lack of identifiability due to label switching i.e. a reordering of indices that can lead to same joint distribution, should be taken into account during the implementation of the algorithm and the interpretation of the corresponding results \cite{Leos2}.

As defined in {Sect.} \ref{sec:hmm1}, the production state space is assumed to comprise $N$ possible values, modelled as a categorical distribution. 
This implies that for each of the $N$ possible states that a latent variable at time $t$ can take, there is a transition probability from this state to each of the $N$ states at time $t+1$, for a total of $N^2$ transition probabilities.  
Since any one transition probability can be determined once the others are known, there are a total of $N(N-1)$ transition parameters to be identified \cite{ai}.
The same reasoning applies to the internal state space (see {Sect.} \ref{sec:hmm2}), where for $K$ possible states, once any one transition probability has been estimated, there are only $K(K-1)$ transition probabilities to determine.

Since each row of a t.p.m. has to add up to 1, the individual parameters of each row are connected and can be considered as a set of parameters to be estimated jointly.
Analogously, all the entries of the stationary distributions take values in $[0,1]$ and are constrained to sum to unity.
Under such requirement, the elements of each of these ``sets'' were sampled conditional on the previous element being drawn.
The first element was sampled independently of any constraint, whereas the last element of each set was estimated as the difference of 1 and the sum of the other estimates in the same set, according to {Sect.} \ref{sec:par}.

\subsection{Prior Distributions}
\label{sec:prior}

In ecological applications, prior distributions are a convenient means of incorporating expert opinion or information from previous o related studies that would otherwise be ignored \cite{king}.
However, in the absence of any expert prior information, an uninformative prior $Beta(1,1)\equiv Unif(0,1)$ was specified for the proportion of observed zeros for the variable \textit{dive wiggliness}. 
Constraining the means of the covariates to be strictly positive, \textit{Log-Normal(log(100),1)} distributions were taken as priors for their estimation, whereas \textit{Inv-Gamma($10^{-3}, 10^{-3})$} were considered for their standard deviations.

Different prior distributions were specified for the transition matrix parameters.
All the entries of the transition probability matrices were assumed to have a $Beta(0.5,0.5)$ prior.

\subsection{Implementation}
The initial values for the estimation of parameters by the MH algorithm were obtained rounding to one decimal place the maximum log-likelihood estimates computed via direct numerical likelihood maximization (using the R function \textsc{nlm}, as in \cite{Leos1}).
The transition probability matrices and the corresponding stationary distributions were kept fixed to their approximated values.

\begin{table}[h!]
\caption{Values obtained for the standard deviations $\delta$ of the Random Walk for each parameter after pilot tuning on 6,000 iterations. The acceptance ratio is reported between parentheses.}
\label{tab:delta}
\centering
\begin{tabular}{lccc}
\toprule
\textbf{\Large Parameter} & \textbf{\large Production} & \textbf{\large Production} &
\textbf{\large Production}\\
& \textbf{\large State 1} & \textbf{\large State 2} & \textbf{\large State 3}\\
\hline \hline
\small
\textsc{Dive Duration} \\
\textit{Mean} & 0.218 (36.36\%) & 0.878 (32.99\%) & 2.200 (33.87\%) \\
\textit{Variance} & 0.258 (32.26\%) & 0.779 (31.40\%) & 1.949 (31.37\%)\\
\hline
\textsc{Maximum Depth} \\
\textit{Mean} & 0.110 (32.82\%) & 0.286 (36.09\%) & 0.972 (32.57\%) \\
\textit{Variance} & 0.102 (32.60\%) & 0.278 (32.52\%) & 0.857 (30.48\%)\\
\hline
\textsc{Dive Wiggliness} \\
\textit{Mean} & 0.081 (32.05\%) & 0.330 (36.93\%) & 1.210 (34.22\%) \\
\textit{Variance} & 0.073 (36.63\%) & 0.320 (36.32\%) & 1.187 (31.65\%)\\
\textit{Zero count} & 0.052 (28.58\%) & 0.008 (29.92\%) & 0.001 (32.30\%) \\
\hline
\end{tabular}
\end{table}

Proposal variances throughout the analyses we tuned adaptively to give resulting acceptance rates of between (see Tab. \ref{tab:delta}) 2\% and 40\% , \cite{gelman,roberts}. 
In this project, an \textit{acceptance fraction} between 25\% and 40\% was deemed appropriate, where the acceptance fraction is the simplest heuristic proxy statistic for tuning, defined as the ratio between the accepted proposed moves over the total number of iterations.
Such range was considered throughout the implementation of all the different versions of the MH algorithm. The procedure was automated in order to perform an adjustment on the standard deviations $\delta$ of the RW every 100 iterations during burn-in and were then fixed for the RW updates of all the following iterations.

To further avoid correlations between parameters, two different approaches to block updating were devised for the MH algorithm.
In the first case, the blocks were built bringing together the parameters of each variable across the three production states in order to update them simultaneously state-wise. 
For each covariate, the means were updated first, followed by the corresponding standard deviations.
Conversely, in the second scheme, parameters were still updated simultaneously state-wise, but the way blocks were arranged changed. 
The means for all the variables were updated first, and then the standard deviations were considered in the same order.
In both cases, the proportions of null values observed for the variable \textit{maximum depth} were studied separately, by simultaneous update.

For the parallel tempering algorithm, a Markov chain with a temperature value $\beta=1$ was used to sample the true energy landscape, while three higher temperature chains with values of $\beta$ equal to 0.75, 0.50 and 0.25 were employed to sample shallower landscapes.
The acceptance probability was based on the difference in energy between consecutive iterations, scaled by the chain temperature, i.e. $min(1, e^{-\beta_j \Delta E_j})$, for $j=1,...,4$.
Every $T=100$ iterations a swap to the higher temperature chains was attempted, for a total of $S=160$ swaps. Pilot tuning was performed on the first 6,000 iterations.
The \textsc{foreach} and \textsc{doParallel} packages in R were used to parallelise such algorithm by allocating it to different cores.
As many clusters as the number of desired temperatures, and thus chains, were created and each allocated to a core, bearing in mind that it is recommended not to use all the available cores on the computer.

\section{Results}

\begin{figure}[!h]
    \centering
    \includegraphics[width=\linewidth]{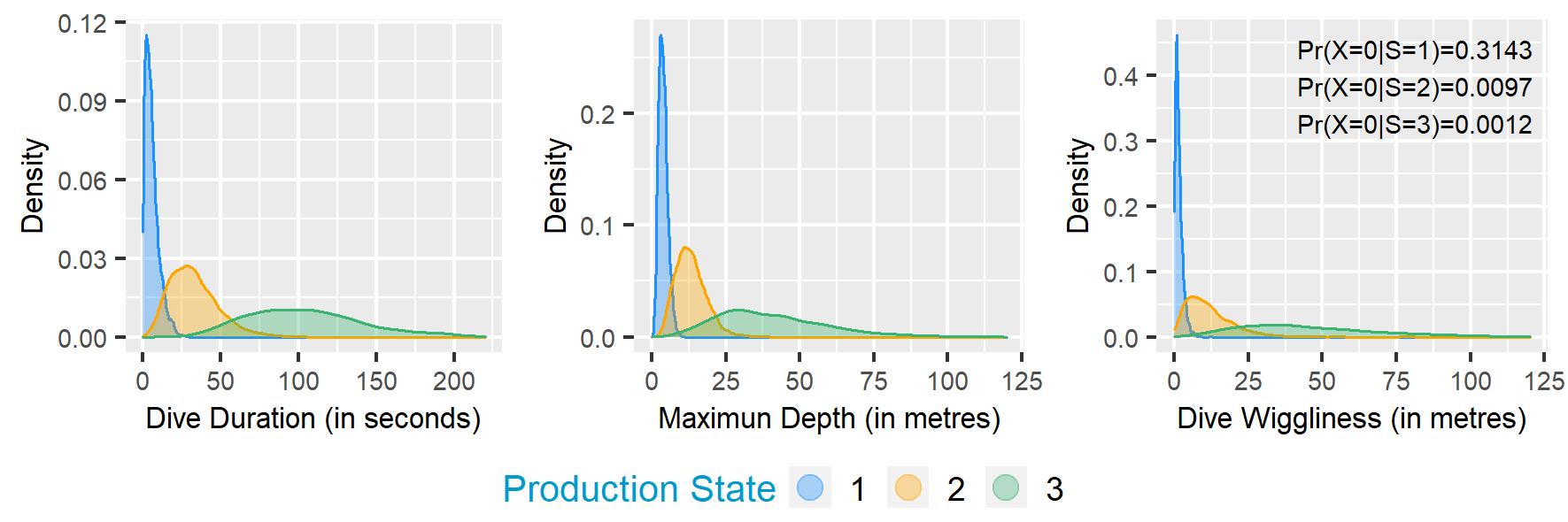}
    \caption{Fitted state-dependent distributions for the variables \textit{dive duration}, \textit{maximum depth} and \textit{dive wiggliness}}
\label{fig:distr}
\end{figure}

Employing the single-update MH algorithm, the fitted (dive-level) state-dependent distributions displayed in Fig. \ref{fig:distr} suggest three distinct dive types. Thresholds for the interpretation of covariates in the three production states were identified as the $2^{nd}$ and the $98^{th}$ percentiles of the corresponding distribution.

\begin{enumerate}
    \item \textit{Production state 1} captures the shortest (lasting less than 20.6s), shallowest (between 1.2 and 7.9m deep) and smoothest (less than 5.7m absolute vertical distance covered) dives with small variance;
    \item \textit{Production state 2} captures moderately long (7.6-77.5s), moderately deep (4.1-28.5m) and moderately wiggly (1–36.3m) dives with moderate variance;
    \item \textit{Production state 3} captures the longest (38–215.7s), deepest (9.7–93.4m) and wiggliest (8.2–120.7m) dives with high variance.
\end{enumerate}

\begin{figure}[h!]
    \centering
    \includegraphics[width=0.96\linewidth]{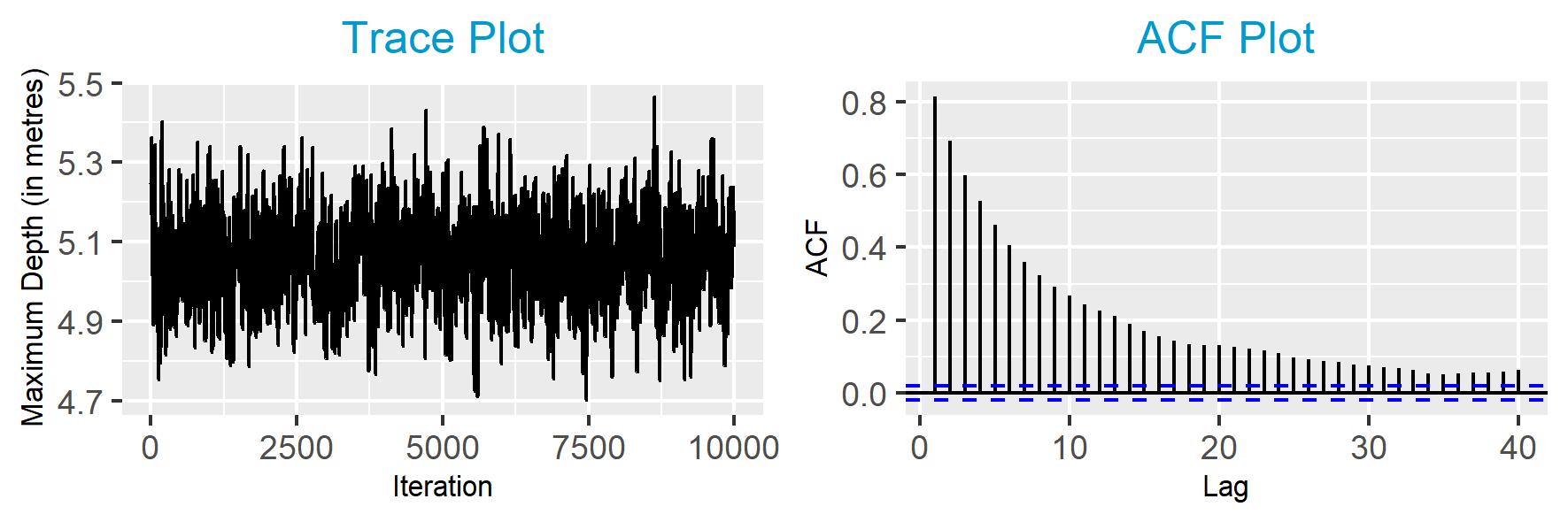}
    \caption{Single-update MH algorithm: trace plot and ACF plot of the standard deviation of the variable \textit{maximum depth} for production state 2 (10,000 iterations)}
\label{fig:acfmh}
\end{figure}

\begin{figure}[!t]
    \centering
    \hspace*{\fill}
    \includegraphics[width=\linewidth]{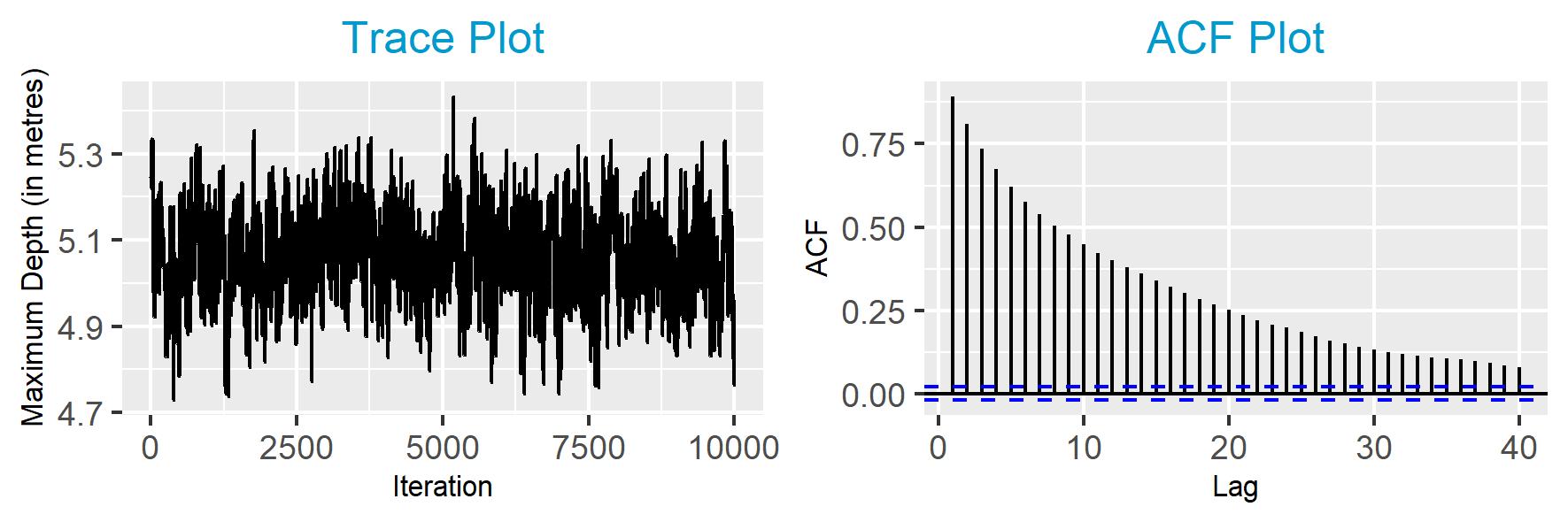}
\hspace*{\fill}
    \includegraphics[width=1\linewidth]{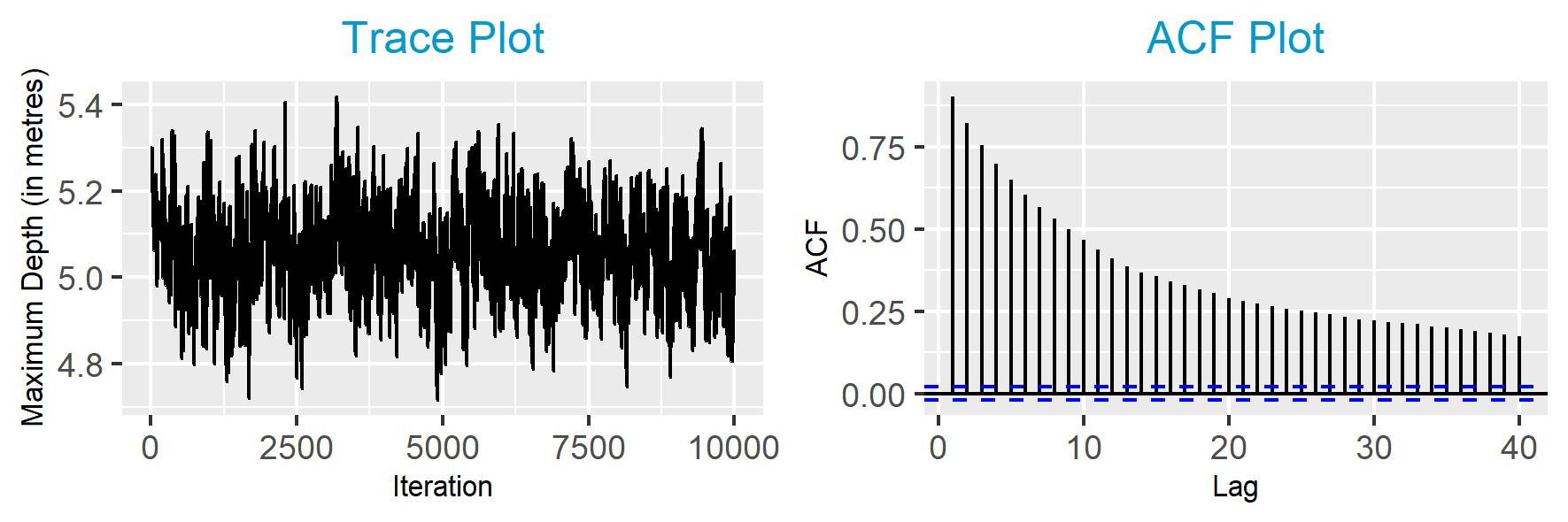}
    \caption{Trace plots and ACF plots of the standard deviation of the \textit{maximum depth} for production state 2 for the two block-update MH algorithms (10,000 iterations). Bwith blocks arranged by variable (upper plots) and parameter (lower plots).}
\label{fig:bl}
\end{figure}

From visual inspection of the trace plots of both the block-update MH algorithms, the first approach appeared to have slightly less autocorrelated chains with better mixing properties, allowing a broader traverse of the parameter space (see Fig. \ref{fig:bl} for a specimen).

The fact that a change in the autocorrelation values happened when altering the order of the blocks might suggest that it could have been worth it to reconsider the original single-update algorithm, varying the order in which parameters were updated.
As a consequence, both algorithms were modified accordingly, relaxing the constraint of simultaneous update for the three production states. 
However, such strategy did not seem to perform much differently from the original MH algorithm nor from the respective block-wise update ones (see Supporting information for corresponding plots).
In particular, multi-parameter algorithms performed slightly worse than the corresponding single-update versions. 
This could be due to the fact that, by updating multiple parameter at the same time during block-wise sampling, a move might have been rejected through being ``poor'' for some parameters, whilst it might have been ``good'' for the others \cite{king}.

It should also be noticed that for multiple parameter moves, the proposed changes are relatively small compared with the current values, as the proposal variance $\delta$ is smaller than for single-parameter updates, leading to a narrower exploration of the support of $h$ \cite{king}.

\begin{figure}[h!]
    \centering
    \includegraphics[width=0.96\linewidth]{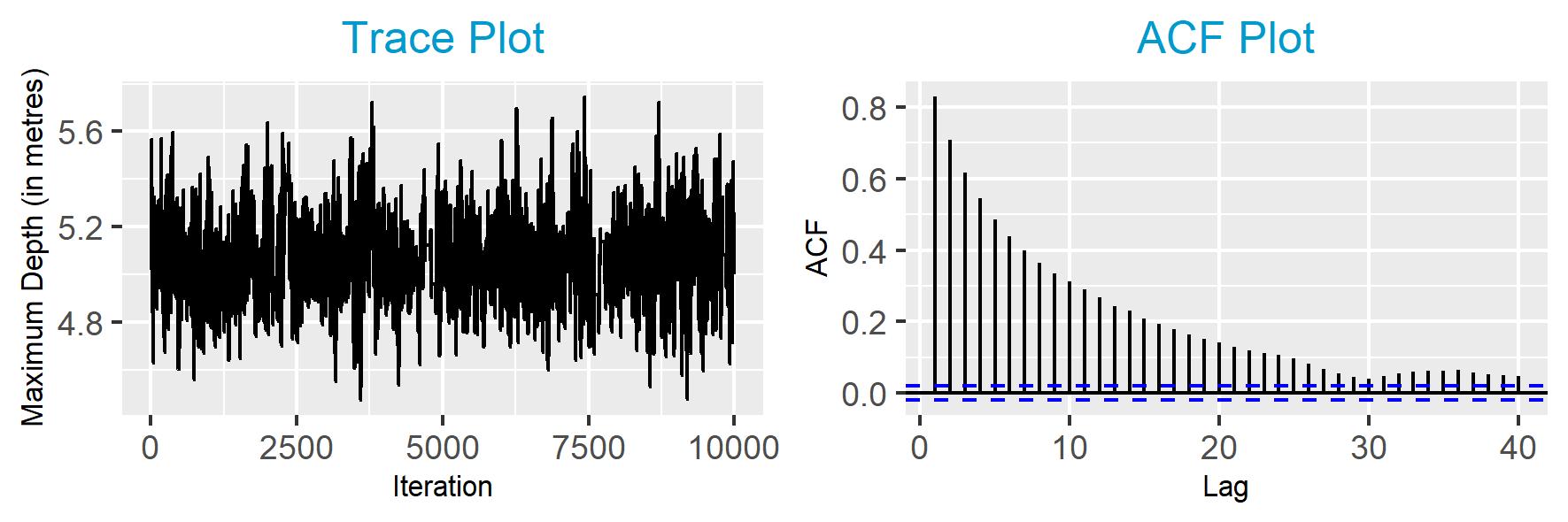}
    \caption{Trace plot and ACF plot of the standard deviation of the variable \textit{maximum depth} for production state 2, obtained running 4 chains for parallel tempering on the MH algorithm (10,000 iterations)}
\label{fig:pt4}
\end{figure}

\subsection{Parallel Tempering}
\label{sec:pt2}

Looking at the trace plots and the ACF plots for the implemented MH algorithm, this strategy performs in a similar way to the single-update MH algorithm, with a good mixing in the chain and a slowly decreasing autocorrelation in the values.  
Fig. \ref{fig:pt4} shows the behaviour of the standard deviation of the variable \textit{maximum depth} for production state 2, when performing parallel tempering with 4 temperature chains.

Due to the high computational effort required by such an alternative, a different strategy was adopted to implement the MH algorithm with parallel tempering. 

\begin{figure}[h!]
    \centering
    \includegraphics[width=0.96\linewidth]{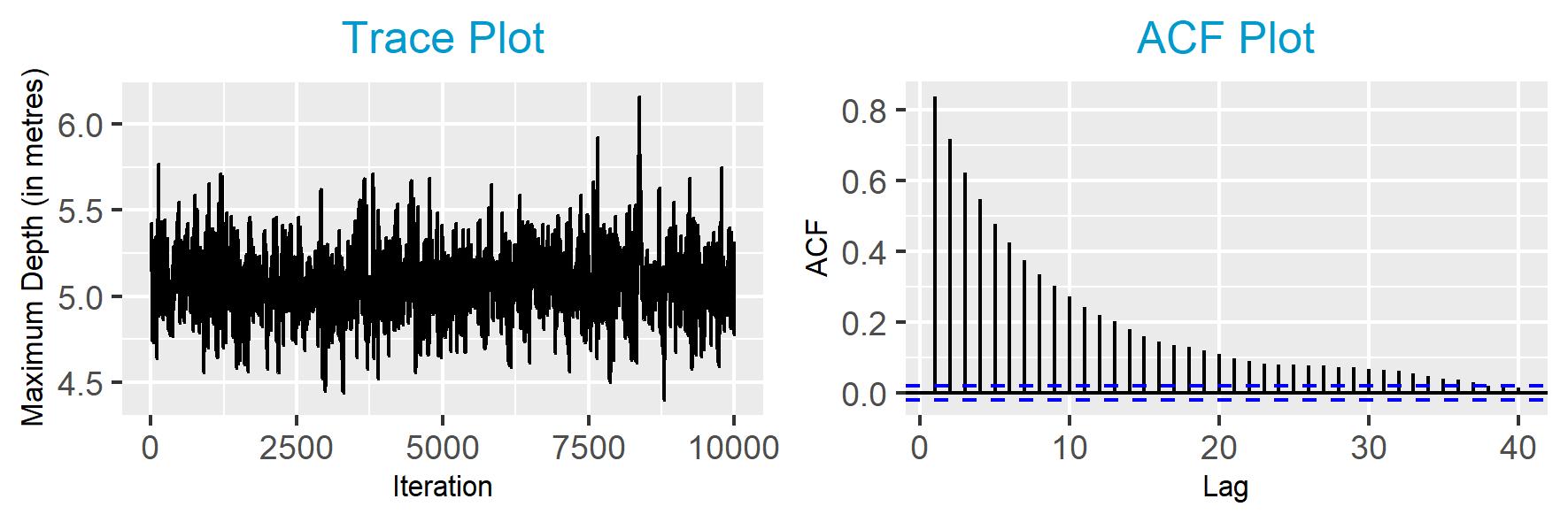}
    \caption{Trace plot and ACF plot of the standard deviation of the variable \textit{maximum depth} for production state 2, obtained running 7 chains for parallel tempering on the MH algorithm (10,000 iterations)}
\label{fig:pt7}
\end{figure}

Considering again the standard deviation of the variable \textit{maximum depth} for production state 2 (see Fig.\ref{fig:pt7}), it can be observed that employing a larger number of temperature chains for parallel tempering generates a better mixing and a roughly smaller autocorrelation between consecutive values.
Nonetheless, the performance of the MH algorithm with parallel tempering varies only to a limited extent with respect to the original MH algorithm.

\subsection{Prior Sensitivity Analysis}

Prior sensitivity analysis was performed on the hyperparameters of the Beta distributions for the proportion of observed null values for the variable \textit{dive wiggliness}.
The MH algorithm with parallel tempering was rerun with Beta priors having both shape parameters either higher ($\alpha=\beta=2$) or lower ($\alpha=\beta=0.5$) than those specified in {Sect.} \ref{sec:prior}.
There was no evidence of significant changes in parameter estimates under varying priors, nor in the performance of the algorithm itself, as reflected in the almost unaffected acceptance rate of proposed moves.

\subsection{Estimation of Transition Probabilities}

In order to proceed with the estimation of the complete collection of parameters, one of the MH algorithms previously implemented was extended to estimate the transition probability matrices and the corresponding stationary distributions.
The original single-update MH algorithm was selected, since it was the one that appeared to perform best.

\begin{figure}[h!]
    \centering
    \includegraphics[width=0.96\linewidth]{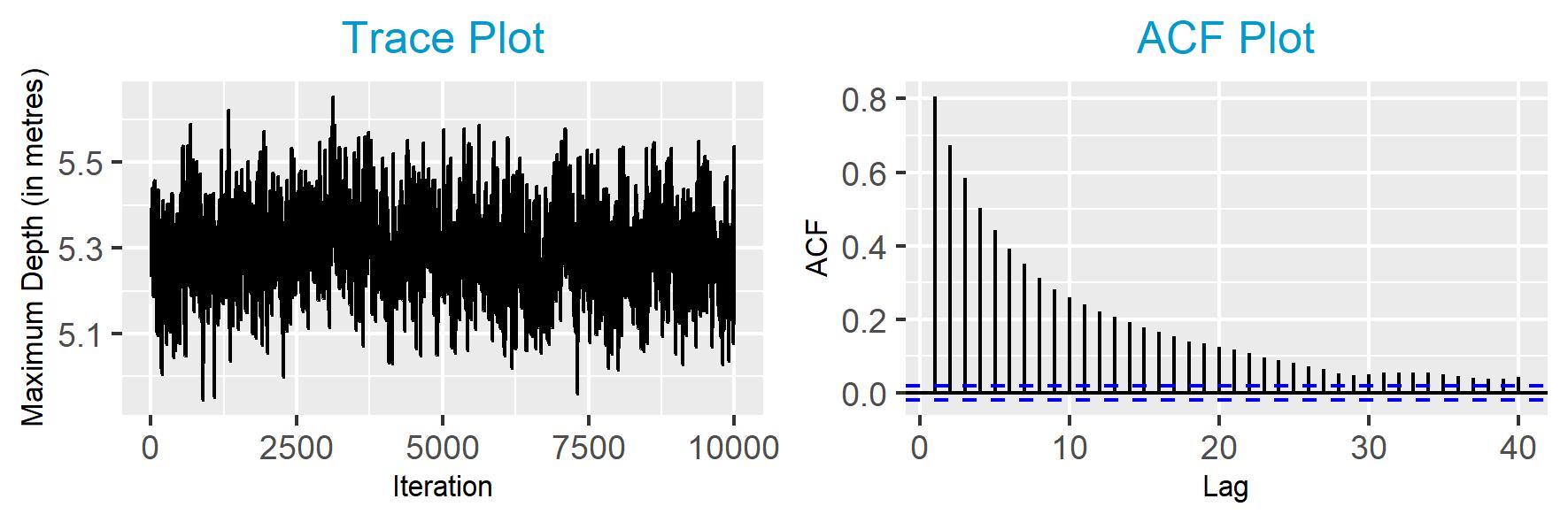}
    \caption{Single-update MH algorithm for the complete parameter set: trace plot and ACF plot of the standard deviation of the variable \textit{maximum depth} for production state 2 (10,000 iterations)}
\label{fig:mhtot}
\end{figure}

Fig.\ref{fig:mhtot} illustrates the behaviour of the standard deviation of the variable \textit{maximum depth}.
It shows good mixing speed and the gradient of the autocorrelation function is not excessively shallow.
A similar behaviour can be observed for most of the other parameters (see Supplementary material).

However, it should be also mentioned that the acceptance fraction for the transition probabilities appeared to be quite high, ranging from nearly 36\% to roughly 90\% (see Supporting information).

\section{Discussion}

According to the ACF plots for all the variables of the implemented MH algorithm, the means and the standard deviations present relatively high autocorrelation, even for larger values of $l$.
In principle, such an issue was partially handled during the pilot tuning, by adjusting the standard deviation $\delta$ of the RW for the parameters of interest.
When the chosen values of $\delta$ were too small, making the proposal distribution very tight, only small moves from the current value were generally accepted with a high probability.
This means that although the probability of accepting a proposed value was high in the MH step, the distance travelled was small. 
Alternatively, when excessively large values of $\delta$ were defined, the high levels of autocorrelation were a result of a small acceptance probability in the MH step. 
Indeed, the parameter often remained at the same value without being updated to a new one for various following iterations, resulting in only a small decrease in the autocorrelation function for increasing lags.

However, pilot tuning proved itself inadequate to reduce the autocorrelation sufficiently.
In order to improve the performance of the algorithm, different strategies were considered, such as block updates and parallel tempering.

Initially, the focus was on estimating the parameters of the gamma distributions of the covariates for all production states, with an additional point mass on zero for the \textit{dive wiggliness} to account for null observations.

At first, a single-update Metropolis-Hastings algorithm was constructed. 
For some parameters, the acceptance fraction fell outside the chosen thresholds, causing a fairly slow decrease in their autocorrelation functions and a non-optimal mixing speed for the corresponding chains. 
Thus, pilot tuning was conducted on the standard deviations of the random walk used for the proposal distributions, resulting in an overall improvement of the performance, yet insufficient to yield a significant reduction in the autocorrelation.

Conjecturing the presence of correlation between parameters, block-wise sampling was employed for simultaneous updates across the three production states.
Two different ways of ordering such blocks were considered. 
In the first case, the means for the three production states of all the covariates were updated first, each followed by the corresponding standard deviations. 
In the second, all the means were updated first, followed by all the standard deviations, according to the same order in the covariates. 
Although the first approach appeared to have less autocorrelated chains with better mixing properties than the second, the global performance turned out not to be better than the single-update one. 
In fact, the simultaneous update of multiple parameters might have caused the rejection of some moves through being inadequate for some parameters, whilst it might have been suitable for the others. 
Moreover, having a closer look at the trace plots and the ACF plots, a modest difference in their autocorrelation values was visible, hinting that the order in which parameters were updated might have had some influence on the performance. 
Thus, both algorithms were modified accordingly to construct two single-update MH algorithms. 
Such strategy did not yield results dissimilar to the original MH algorithm, indicating that the order in which parameters were arranged did not have a significant influence on the behaviour of the chain.

In many cases it was necessary to make relatively small moves for the parameters to be accepted, leading to a narrower exploration of the support of the target distribution.
For other parameters, instead, the acceptance rate turned out to be especially high.
Indeed, the selection of non-ideal values for the standard deviations $\delta$ was mirrored in sub-optimal behaviour of the chains, underlining a strong dependence on the choice of the proposal standard deviations.
In fact, even if the chains exhibited good mixing speed, the autocorrelation values were still quite high even for large lags, especially for the production state 1.
This suggested that an appropriate implementation of pilot tuning might have been instrumental in ameliorating the overall performance.

Therefore, a new MH algorithm involving parallel tempering was implemented. 
Different chains were run in parallel, each linked to an auxiliary temperature value that enabled to sample from a distinct energy landscape. 
At regular intervals, the behaviour of the different chains was analysed and a swap between temperature chains allowed, according to an acceptance probability that depended on the difference in both temperature and energy between different temperature chains. 
This facilitated the traverse of the sample space, exploiting the concept that higher temperature chains explore the sample space more broadly and are able to circumvent local minima. 
In the light of the fact that distributions on neighbouring temperature levels need to have a considerable overlap, the number of temperatures was increased from four to seven.
Running several chains from distinct points, as in parallel tempering, ensures the exploration of multimodal distributions.
In fact, the posterior density plots for parallel tempering were globally unimodal, whereas for most of the other algorithms, many distributions showed some gentle spikes (see Supporting information).
However, this strategy seemed to perform in a similar way to the single-update MH algorithm, with good mixing in the chain and a fairly high autocorrelation even for quite large lags.
In particular, the performance of the MH algorithm with more temperature chains improved only to a limited extent with respect to the one with fewer chains. In practice, to have a reasonable acceptance rate, the temperatures need to be chosen carefully, and checked through preliminary runs. 
Moreover, the distributions on neighbouring temperature levels need to have a considerable overlap \cite{hogg}, suggesting that it would be appropriate to consider temperature values which are relatively close and possibly evenly spaced.
For this purpose, the energy landscape was subsequently explored with a larger number $J$ of chains, with temperatures $\beta$ taking values equal to 1, 0.857, 0.714, 0.571, 0.429, 0.286, 0.143.

\begin{table}[h!]
\caption{Parameter estimates for the three production states, found through Maximum Likelihood Estimation and Bayesian Inference (on the complete collection of parameters)}
\label{tab:est_tot}
\centering
\begin{tabular}{lrrrrrr}
\toprule
\multicolumn{1}{l}{}& \multicolumn{3}{c}{\textbf{\Large{MLE}}} & \multicolumn{3}{c}{\textbf{\Large{Bayesian Estimate}}}\\
\textbf{\Large{Parameter}} & \textbf{\normalsize{Prod.}} & \textbf{ \normalsize{Prod.}} & \textbf{\normalsize{Prod.}}  & \textbf{\normalsize{Prod.}} & \textbf{ \normalsize{Prod.}} & \textbf{\normalsize{Prod.}}\\
& \textbf{\normalsize{State 1}} & \textbf{\normalsize{State 2}} & \textbf{\normalsize{State 3}} & \textbf{\normalsize{State 1}} & \textbf{\normalsize{State 2}} & \textbf{\normalsize{State 3}}\\
\hline \hline
\textsc{Dive Duration} \\
\textit{Mean} & 5.633 & 32.167 & 106.817 & 5.711 & 32.321 & 106.891 \\
\textit{Variance} & 4.368 & 15.138 & 38.417 & 4.440 & 15.159 & 38.391\\
\hline
\textsc{Maximum Depth} \\
\textit{Mean} & 3.738 & 13.188 & 39.613 & 3.766 & 13.241 & 39.658 \\
\textit{Variance} & 1.425 & 5.288 & 18.111 & 1.449 & 5.293 & 18.135\\
\hline
\textsc{Dive Wiggliness} \\
\textit{Mean} & 1.455 & 11.292 & 45.822 & 1.482 & 11.362 & 45.893\\
\textit{Variance} & 1.203 & 7.681 & 24.374 & 1.233 & 7.701 & 24.402\\
\textit{Zero count} & 0.309 & 0.008 & 0.0002 & 0.307 & 0.008 & 0.0006\\
\hline
\end{tabular}
\end{table}

\subsection{Comparison of Frequentist and Bayesian Results}

The estimates found through Bayesian inference did not differ considerably from the ones obtained in \cite{Leos1} using maximum likelihood estimation (Tab. \ref{tab:est_tot}).
This does not alter significantly the understanding of the ecological process, suggesting that the two methods might be equally effective in terms of interpreting the behaviour of a harbour porpoise.

The first HMM might be indicative of a foraging behaviour, particularly due to the extensive wiggliness characterising the dives, which often indicates prey-chasing. 
Conversely, the second HMM could represent a resting and/or a travelling behaviour, due to the prevalence of relatively short, shallow and smooth dives. 

It is interesting to notice that the estimated t.p.m.s for the upper levels and the corresponding stationary distributions, though leading to an analogous behavioural interpretation, appear in reverse order for the two estimation methods.
This allows to remark that Hidden Markov Models are mixture models, and hence they are invariant under permutations of the indices of the components \cite{abm}, meaning that the naming of the internal and production states does not correspond to a given natural ordering.

Furthermore, it could be worthwhile to look at the estimates returned by the single-update MH algorithm, both when the t.p.m.s and their invariant distributions were kept constant (see Supporting information) and when instead they were included in the estimation process.
Although one might expect the estimates for the first algorithm to be closer to the MLEs, as the hierarchical structure is more stable, the ones from the second algorithm are actually more similar to the frequentist results.
In fact, while in the first case the transition probabilities and the stationary distributions are kept fixed to a very loose approximation of the classical estimates, in the second they will gradually converge to their actual estimates, easing the convergence of the whole set.

The estimates of the single-update algorithm for the partial set of parameters could be also be compared to the ones obtained when employing parallel tempering.
Even if they are roughly the same, the second algorithm is slightly closer to the values for the complete collection of parameters, using both classical and Bayesian inference.
This is linked to the fact that employing parallel tempering enhances the sampler, by handling multimodality and conducting a wider exploration of the sample space.

\section{Conclusions}

The implemented algorithms appear to perform satisfactorily and to provide a sensible insight on behavioural modes connected to the analysed movement data.

In such a way, a more robust structure might be constructed for block sampling, by specifying a suitable multi-dimensional proposal distribution, as recommended by \cite{king}. 
Doing so,a more efficient strategy for pilot tuning could be devised as well.
For instance, following the suggestions provided by \cite{delta}, the adjustment of the standard deviations $\delta$ of the RW updates could be performed automatically also for the transition probabilities.
If possible, it would also be advisable to perform parallel tempering on the complete collection of parameters, or at least run multiple chains starting from overdispersed points in order to study the convergence of the chains.

From a broader perspective, Metropolis-Hastings algorithms might show an inherent inefficiency linked to the random walk behaviour. As in this study, such features may induce a lengthy exploration, requiring a larger number of iterations as the dimension of the problem increases and the complexity of the data behind it grows.
\cite{acc} illustrates several techniques for overcoming such issues and accelerating the convergence of MCMC algorithms.

Another important issue that may arise in the context of movement modelling is that the computational effort required to calculate the likelihood might be noticeably large compared to that of the simulation for the model of interest.
In fact, for complex models - such as hierarchical HMMs - deriving an analytical formula for the likelihood function might be elusive or computationally expensive.
\textit{Approximate Bayesian Computation} methods are simulation-based techniques useful to infer parameters and choose between models, bypassing the evaluation of the likelihood function \cite{beaumont, abc}.
Following the results of \cite{ruiz}, for this particular stochastic problem it could be worthwhile to employ a likelihood-free method based on the measure of similarity between simulated and actual observations.
In fact, while the likelihood function is quite intractable due to the complicated relationship between parameters, simulating the movement of an harbour porpoise appears to be more straightforward.
Therefore, through the simulation of dives from the HMM, relying on independent draws combined with data observed at regular time intervals, it would be possible to easily estimate the posterior distributions of model parameters \cite{abc}.

Bayesian inference has proven to be a valid statistical instrument for modelling animal movement data according to a hierarchical HMM, providing an effective framework to infer drivers of variation in movement patterns, and thus describe distinct behaviours.

Although the computational times related to Bayesian inference tend to be larger than for the frequentist context, working in such a framework provides tools to circumvent issues frequent in statistical ecology and widens the opportunities for further analysis.
For example, while there is no guarantee that a maximum likelihood procedure is really finding all the modes in a distribution, some Bayesian techniques, such as parallel tempering, enable to identify and handle multimodality, which is a common problem in mixture models and in the field of ecology.
Another prerogative of the Bayesian framework is incorporating in the MCMC algorithm biological a priori knowledge, which might provide a considerable benefit in terms of the estimation.

Keeping in mind that different algorithms and computational methods are usually complementary - not competitive, the tools available to the Bayesian paradigm provide a very powerful and flexible framework for the analysis of complex high-dimensional stochastic processes.

%\clearpage

%These commands reset the figure counter and add "S" to the figure caption (e.g. "Figure S1"). This is in case you want to add actual figures and not just captions.
\setcounter{figure}{0}
\renewcommand{\thefigure}{S\arabic{figure}}

% You can use the \nameref{label} command to cite supporting items in the text.
%\clearpage

\nolinenumbers

%This is where your bibliography is generated. Make sure that your .bib file is actually called library.bib

%This defines the bibliographies style. Search online for a list of available styles.
\bibliographystyle{abbrv}

\end{document}